\documentclass[twocolumn,showpacs,prb,amsmath,amssymb,floatfix]{revtex4}
\usepackage{amsmath,amssymb}
\usepackage{bm}
\usepackage{epsfig}
\usepackage{psfrag}

\newcommand{\bd}{\bm}

\begin{document}

\title{Effective spin-wave action for ordered 
Heisenberg antiferromagnets in a magnetic field}

\author{Nils Hasselmann, Florian Sch\"{u}tz, Ivan Spremo, and Peter Kopietz}  
\affiliation{Institut f\"{u}r Theoretische Physik, Universit\"{a}t
  Frankfurt, Max-von-Laue-Strasse 1, 60438 Frankfurt, Germany}

\date{November 29, 2005}

\begin{abstract}
  We derive the effective long-wavelength Euclidean action for the
  antiferromagnetic spin-waves of ordered quantum antiferromagnets
  subject to a uniform magnetic field.  We point out that the magnetic
  field dependence of the spin-wave dispersion predicted by the usual
  $O (3)$-quantum nonlinear sigma model disagrees with spin-wave theory.
  We argue that the nonlinear sigma model does not take
  into account all relevant spin-wave interactions and derive a
  modified effective action for the long-wavelength spin-waves which
  contains an additional quartic interaction. At zero temperature the
  corresponding vertex is relevant in the renormalization group sense
  below three dimensions.

\end{abstract}




\maketitle


\section{Introduction}

Recently, two of us derived a new effective field theory\cite{Hasselmann05} 
for transverse spin fluctuations in ordered quantum 
antiferromagnets. 
Our approach is based on 
the Holstein-Primakoff transformation,\cite{Holstein40} 
which maps the original
spin Hamiltonian onto a bosonic many-body Hamiltonian. However,
rather than working directly with the canonical 
Holstein-Primakoff bosons,
we express them
in terms of Hermitian field operators representing staggered and uniform
transverse spin fluctuations. \cite{Anderson52}
The main advantages of this procedure
are: (a) the suppression of the effective interaction between
antiferromagnetic magnons at long wavelengths 
is manifest; (b) the relation between
Holstein-Primakoff bosons and the fields of the
nonlinear sigma model\cite{Chakravarty89} (NLSM) representing
the transverse staggered spin fluctuation is made precise; 
(c)  the
zero wave vector modes of the uniform and staggered 
magnetization can be easily treated within this approach,\cite{Anderson52} 
which is important in the study of finite magnets;
(d) the transverse ferromagnetic fluctuations
may be eliminated to yield an effective theory involving only transverse
antiferromagnetic fluctuations.
This is most easily done within a
path integral approach; the resulting
effective action coincides with the NLSM
at the Gaussian level. 
Although the interaction terms  of the effective action differ from
those of the NLSM,
the renormalization group flows of both theories agree to 
one-loop order.\cite{Hasselmann05} 

Here, we extend the Hermitian operator approach to antiferromagnets
in a uniform magnetic field. In this case, the differences
between this approach and the NLSM become more substantial. 
Already at the Gaussian level the two theories differ: while
the Hermitian operator approach reproduces by construction
the correct
spin-wave dispersion known from spin-wave theory, the NLSM
does not. More severely,
we show below that a quartic interaction term present
in the Hermitian operator approach is absent in the NLSM. 

We shall constrain the analysis here to a Heisenberg model on a
hypercubic lattice
with nearest neighbor antiferromagnetic coupling $J>0$.
It is however straightforward to extend the results to general
bipartite lattices. This would also cover a
recently synthesized metal-organic $S =5/2$ quantum antiferromagnet on a
distorted honeycomb lattice,\cite{Spremo05} which is bipartite.
In presence of a uniform external field $\bd h$ (measured
in units of energy), the Hamiltonian is
given by
\begin{equation}
  \hat{H} = 
J  \sum_{\left<ij\right>}\bd{S}_i  \cdot \bd{S}_j  -
  \sum_i \bd{h} \cdot 
  \bd{S}_i 
  \;.
  \label{eq:hamiltonian}
\end{equation}
Here, $\bd{S}_i$ are spin operators normalized such that $\bd{S}_i^2 =
S ( S+1)$, 
and the first sum is over nearest neighbor pairs $\left<ij\right>$ of a
hypercubic lattice with lattice constant $a$. 

Our main
interest is a 
description of the low-energy and long-wavelength 
properties of the Heisenberg model (\ref{eq:hamiltonian}). 
A standard approach to this problem is the $O(3)$ quantum NLSM whose
effective action describes only  the
relevant staggered spin fluctuations. 
According to Refs.~[\onlinecite{Sachdev99,Fisher89}], a magnetic
field can be taken into account within the NLSM approach 
by
replacing the derivative $\partial_{\tau}$ with respect to the
imaginary time $\tau$
with a covariant derivative (or Lie
derivative, see Ref.~[\onlinecite{McCauley97}])
\begin{equation}
  \partial_{\tau}  \to \partial_{\tau}  - i \bd{h} \times
  \; .
  \label{eq:minimal}
\end{equation}
With this minimal coupling the NLSM in a uniform
magnetic field has the form
\begin{eqnarray}
  S_{\rm{NLSM} }
  [ \bd{\Omega} ] 
  & = & \frac{ \rho_0}{2} \int_0^{\beta} d \tau
  \int d^{D} r \Bigl[ \sum_{ \mu =1}^{D} ( \partial_{\mu} \bd{\Omega})^2 
  \nonumber
  \\
  & & \hspace{10mm}
  + \frac{1}{c_0^2} ( \partial_{\tau}  \bd{\Omega}  - i \bd{h} \times  
  \bd{\Omega})^2 \Bigr]
  \; ,
  \label{eq:sigmah2} 
\end{eqnarray}
where the unit vector $\bd{\Omega} ( \tau , \bd{r} )$ represents the
slowly fluctuating staggered magnetization, $\rho_0$ and $c_0$ are the
spin stiffness and spin-wave velocity at temperature $T = 1/ \beta
=0$, and $\partial_{\mu} =
\partial / \partial r_{\mu}$ is the spatial derivative in direction
$\mu = 1, \ldots , D$.  Numerical values of $\rho_0$ and
$c_0$ must be computed microscopically, e.~g.~by
means of a $1/S$-expansion.\cite{Chakravarty89}  
To leading order in $1/S$, 
one has $\rho_0 \approx JS^2 a^{2-D}$ and $c_0 \approx 2 D^{1/2} JS a$.
For later reference, we note that 
the classical uniform transverse susceptibility is
$ \chi_0 = \rho_0 / c_0^2$ with $\chi_0\approx ( 4 D {J} a^D )^{-1}$
to leading order in $1/S$.
The classical
uniform magnetization per volume is 
$ \bd{M}_0 = \chi_0 \bd{h} $.

\section{Spin-wave dispersion}

\begin{figure}[tb]    
  \centering
  \psfrag{om}{$n_0 \bd{e}_z$}
  \psfrag{n1}{$\hat{\bd{m}}_i$}
  \psfrag{n2}{$\hat{\bd{m}}_j$}
  \psfrag{m}{$\bd{m}_0$}
  \psfrag{oo}{$- n_0 \bd{e}_z$  }
  \psfrag{t}{$\vartheta_0$}
  \psfrag{e2a}{$\bd{e}_A^2$}
  \psfrag{e2b}{$\bd{e}_B^2$}
  \psfrag{B}{$\bd{h}$}
  \vspace{7mm}
  \epsfig{file=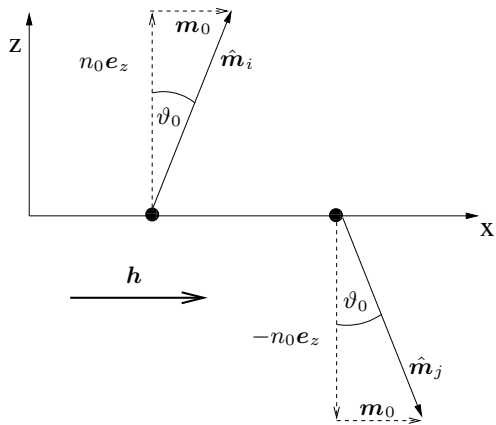,width=65mm}
  \vspace{4mm}
  \caption{%
    Spin configuration $\langle \bd{S}_i \rangle  = S \hat{\bd{m}_i}$ 
    in the classical ground state
    of a two-sublattice antiferromagnet subject to a uniform magnetic field
    $\bd{h} = h \bd{e}_x$, where $\bd{r}_i$ belongs to sublattice $A$ and 
    $\bd{r}_j$ to sublattice $B$.
    Here ${\bd m}_0 = \bd{h}  /( 4 D JS )$ and $ n_0 = \sqrt{ 1 - m_0^2}$.
}
\label{fig:spinconfig}
\end{figure}

In absence of a magnetic field, the NLSM 
predicts a doubly degenerate antiferromagnetic spin-wave mode with
long-wavelength dispersion $E_{\bd{k}} = c_0 | {\bd{k} } |$, in
agreement with linear spin-wave
theory.\cite{Chakravarty89} 
Rather surprisingly, however, in presence of a magnetic field 
the long-wavelength dispersion
derived from the NLSM 
in Eq.~(\ref{eq:sigmah2}) disagrees with  spin-wave theory.  
To see this, let us recall the
spin-wave dispersion predicted by linear spin-wave theory.  In this
approach the spin operators are expanded in deviations from the 
classical ground state spin configuration, using either the
Holstein-Primakoff \cite{Holstein40} or the Dyson-Maleev\cite{Dyson56}
transformation.  The classical ground state of a quantum
antiferromagnet in a uniform external field is canted, 
where the staggered magnetization is perpendicular to
the magnetic field, as shown in Fig.~\ref{fig:spinconfig}.
To leading order in
$1/S$, spin-wave
 theory\cite{Zhitomirsky98,Zhitomirsky99,Spremo05} predicts
two transverse spin-wave modes,
one gapless and one gapped, with long-wavelength dispersions 
(see Eqs.~(\ref{eq:epsilondef},\ref{eq:Edispersiondef}) below)
\begin{equation}
  E_{ \bd{k}  +} \approx  [ h^2 + c_+^2   \bd{k}^2  ]^{1/2}
 \; , \; \;
  E_{ \bd{k}  -} \approx c_- | {\bd{k}} |
  \; ,
  \label{eq:dispersion}
\end{equation}
where the spin-wave velocities are
\begin{equation}
  c_+ = c_0 \sqrt{ 1 - 3 m_0^2} \; , \; c_- = c_0 \sqrt{ 1 - m_0^2}
  \label{eq:spinwaveh}
  \; .
\end{equation}
Here, $m_0 = | {\bd{m}}_0 |$ is the length of the normalized classical
uniform magnetization
\begin{equation}
  \bd{m}_0 = a^D \frac{ \bd{M}_0 }{ S} = \frac{ \bd{h}}{4 D J  S } 
  \; ,
  \label{eq:m0def}
\end{equation}
which is related to the classical canting angle $\vartheta_0$
between the direction of the
staggered magnetization and the local magnetic moments by
$m_0 = \sin \vartheta_0 $. In general, the classical canting angle 
is renormalized by quantum
fluctuations.\cite{Spremo05,Zhitomirsky98}

Let us compare Eqs.~(\ref{eq:dispersion},\ref{eq:spinwaveh})
with the dispersion relation obtained from
the NLSM. Denoting by
$\bd{e}_{\alpha}$ the unit vectors in the fixed directions $\alpha =
x,y,z$ and assuming $\bd{h}=h \bd{e}_x$ 
to point along the $x$-axis, we
write
\begin{equation}
  \bd{\Omega} = \Pi_{\parallel} \bd{e}_x + \Pi_{\bot} \bd{e}_y +
  [ 1 - \Pi_{\parallel}^2 - \Pi_{\bot}^2 ]^{1/2} \bd{e}_z
  \; ,
\end{equation}
and expand the action $S_{\rm NLSM} [ \bd{\Omega} ]$ to quadratic
order in the fluctuations $\Pi_{\parallel}$ and $\Pi_{\bot}$
transverse to the direction of the staggered magnetization.  One
easily finds from Eq.~(\ref{eq:sigmah2}) that the spin-wave mode
$\Pi_{\parallel}$ polarized parallel to the magnetic field is gapped
with energy dispersion
\begin{equation}
  E_{ \bd{k} \parallel } =  [h^2 + c_0^2 \bd{k}^2 ]^{1/2}
  \; ,
  \label{eq:dispersionparallel}
\end{equation}
while the spin-wave mode $\Pi_{\bot}$ polarized perpendicular to the
magnetic field has a gapless linear dispersion,
\begin{equation}
  E_{ \bd{k} \bot } = c_0   | {\bd{k}} |
  \; .
  \label{eq:dispersionperp}
\end{equation}
  The spin-wave spectrum of the
NLSM given in Eqs.~(\ref{eq:dispersionparallel}) and
(\ref{eq:dispersionperp}) does not reproduce the correct magnetic
field-dependence of the spin-wave spectrum in
Eqs.~(\ref{eq:dispersion},\ref{eq:spinwaveh}) obtained via conventional
spin-wave theory. Obviously, the above minimal coupling
(\ref{eq:minimal}) does not  account
for order $(h/J)^2$ corrections to the dispersion, even though
the qualitative aspects of the dispersion are correctly
described. 

\section{Effective Action}

In Ref.~[\onlinecite{Hasselmann05}] we have developed a new method to derive 
the effective action for staggered transverse
spin-fluctuations.  
We here sketch the main steps of the derivation adapted
to a quantum antiferromagnet subject to 
a uniform magnetic field.\cite{Hasselmann05b} 
Starting point is the representation
of spin operators in terms of canonical boson operators $b_i$ and
$b_{i}^{\dagger}$ using the Holstein-Primakoff transformation\cite{Holstein40}.  
Let us
denote by $\hat{\bd{m}}_i = \langle \bd{S}_i \rangle / S$
the directions of the spins in the true ground
state and introduce a local right-handed orthogonal triad of unit
vectors ${\bd{e}}_i^{(1)}, {\bd{e}}_i^{(2)} , \hat{\bd{m}}_i $.
As explained in Ref.~[\onlinecite{Schuetz03}]
there is a $U(1)$ gauge freedom in the choice of the transverse
vectors ${\bd{e}}_i^{(1)}$ and ${\bd{e}}_i^{(2)}$.
Defining the spherical basis vectors ${\bd{e}}^{p}_i
$ $=$ $ {\bd{e}}^{(1)}_i + i p {\bd{e}}^{(2)}_i$, $p $ $=$ $ \pm$, we
express the components of the spin operator $\bd{S}_i $ in terms of
canonical boson operators $b_i$ and $b_i^{\dagger}$ as
follows,\cite{Holstein40}
\begin{equation}
  {\bd{S}}_i =
  S^{\parallel}_i \hat{\bd{m}}_i + {\bd{S}}^{\bot}_i
  = S^{\parallel}_i \hat{\bd{m}}_i +
  \frac{1}{2} \sum_{ p = \pm } S_i^{-p} {\bd{e}}^{p}_i 
  \; ,
  \label{eq:spinexpansion}
\end{equation}
with
\begin{subequations}
  \begin{eqnarray}
    S_i^{\parallel} & = & S - n_i \; \; , \; \; n_i = b^{\dagger}_i b_i
    \; ,
    \label{eq:HP1}
    \\
    S_i^{+} & = & (2S)^{1/2} \Big( 1 - \frac{n_i}{2S}\Big)^{1/2} b_i
     \; , 
    \label{eq:HP3}
  \end{eqnarray}
\end{subequations}
and $S_i^{-}=\big(S_i^{+}\big)^\dagger$.
The Heisenberg model (\ref{eq:hamiltonian}) can then be written
as a bosonic many-body Hamiltonian.\cite{Spremo05}  
We expand around the classical ground state configuration 
(see Fig.~\ref{fig:spinconfig})
\begin{eqnarray}
  \hat{\bd{m}}_i&=&\zeta_i n_0 \bd{e}_z
  +m_0 \bd{e}_x \; ,
\end{eqnarray}
and choose 
\begin{equation}
  \bd{e}^{(1)}_i=\bd{e}_y \;, \; \; \;
  \bd{e}^{(2)}_i=-\zeta_i n_0\bd{e}_x+m_0\bd{e}_z \; ,
  \label{eq:transverse}
\end{equation}
where $\zeta_i=1$ 
($\zeta_i=-1$)
for $\bd{r}_i\in A$ ($\bd{r}_i\in B$). Here, 
$ n_0 = \sqrt{ 1 - m_0^2}=\cos\vartheta_0$.
To diagonalize the quadratic part of this Hamiltonian, 
we first perform a
Fourier transformation in the sublattice basis 
\begin{equation}
  b_i  =  \left\{
    \begin{array}{ll}
      (2/N)^{1/2}
      \sum_{ \bd{k} } e^{ i \bd{k} \cdot
        \bd{r}_i } A_{\bd{k}} \; ,& {\bd r}_i\in{ A} \; , \\
      (2/N)^{1/2}
      \sum_{ \bd{k} } e^{ i \bd{k} \cdot
        \bd{r}_i } B_{\bd{k}} \; ,& {\bd r}_i \in{ B}\; ,
    \end{array} \right.
  \label{eq:akdef}
\end{equation}
where the momentum sums 
are over the reduced
(antiferromagnetic) Brillouin zone. 
After introducing the symmetric and antisymmetric combinations
$C_{\bd{k} \sigma } = 2^{-1/2} [ {A}_{\bd{k}} +\sigma
B_{\bd{k}} ]$, $\sigma=\pm$,  we apply the 
Bogoliubov transformation
\begin{equation}
  \left( \begin{array}{c}
      C_{ \bd{k} \sigma } \\
      C^{\dagger}_{ - \bd{k} \sigma }  \end{array}
  \right) =
  \left( \begin{array}{cc}
      u_{ \bd{k} \sigma} & - \sigma v_{\bd{k} \sigma} \\
      -  \sigma v_{\bd{k} \sigma} & u_{ \bd{k} \sigma} \end{array} \right)
  \left( \begin{array}{c}
      \hat{\Psi}_{ \bd{k} \sigma } \\
      \hat{\Psi}^{\dagger}_{ - \bd{k} \sigma }  \end{array}
  \right)
  \; ,
  \label{eq:bogoliubov}
\end{equation}
where
\begin{subequations}
  \begin{eqnarray}
    u_{ \bd{k} \sigma } & = & 
    \Big[ \frac{ 1 +\sigma m_0^2 \gamma_{\bd{k}} + \epsilon_{\bd{k} \sigma} }{ 2 \epsilon_{\bd{k} \sigma}} \Big]^{1/2}
    \; ,
    \label{eq:ukdef}
    \\
    v_{ \bd{k} \sigma } & = &
    \Big[ \frac{ 1 +\sigma m_0^2 \gamma_{\bd{k}} - \epsilon_{\bd{k} \sigma} }{ 2 \epsilon_{\bd{k} \sigma}} \Big]^{1/2}
    \; ,
    \label{eq:vkdef}
  \end{eqnarray}
\end{subequations}
with
\begin{equation} 
  \epsilon_{ \bd{k} \sigma }  =   \bigl[
 (  1 +\sigma m_0^2  \gamma_{\bd{k}} )^2 - 
 ( n_0^2 \gamma_{\bd{k}} )^2 \bigr]^{1/2}
 \label{eq:epsilondef}
 \; .
\end{equation}
Here, $\gamma_{\bd k}= D^{-1} \sum_\mu
\cos(   \bd{k} \cdot \bd{a}_\mu  )$, where
${\bd a}_\mu$, $\mu =1 \ldots D$, are the $D$ primitive lattice vectors
of the hypercubic lattice.
The quadratic part of the effective boson
Hamiltonian assumes then the  form of non-interacting harmonic 
oscillators,\cite{Spremo05,Hasselmann05b}
\begin{equation}
  \hat{H}_2 =  \sum_{ \bd{k},\sigma } 
  E_{ \bd{k} \sigma}  \left[ \hat{\Psi}^{\dagger}_{ \bd{k} \sigma}
    \hat{\Psi}_{ \bd{k} \sigma} + \frac{1}{2} \right] 
  \; ,
  \label{eq:H2diagpsi}
\end{equation}
where the operators $\hat{\Psi}_{\bd{k} \sigma}$ satisfy the canonical
bosonic commutation relations,
\begin{equation} 
  [ \hat{\Psi}_{ \bd{k} \sigma} , \hat{\Psi}^{\dagger}_{ \bd{k}^{\prime}  \sigma^{\prime}} ] =
  \delta_{ \bd{k} , \bd{k}^{\prime} } \delta_{ \sigma , \sigma^{\prime} }
  \label{eq:Psicom}
  \; ,
\end{equation}
and the energy dispersion of the two spin-wave modes 
are given by
\begin{equation}
  E_{ \bd{k} \sigma } = 2 D J S \epsilon_{ \bd{k} \sigma }
  \; .
  \label{eq:Edispersiondef}
\end{equation}
 In the limit of long
wavelengths 
Eq.~(\ref{eq:Edispersiondef})
reduces to the energy dispersions given in 
Eqs.~(\ref{eq:dispersion},\ref{eq:spinwaveh}).

To derive the long-wavelength effective  action for the staggered transverse
spin fluctuations, we need the precise relation between the
Bogoliubov quasi-particle operators $\hat{\Psi}_{\bd{k} \sigma}$ and
the field operators $\hat{\Pi}_{ \bd{k} \sigma}$ 
representing the transverse fluctuations of the staggered
magnetization.  In a Euclidean path integral approach, these operators
correspond to continuum fields $\Pi_{ \bd{k} \sigma} ( \tau )$
which are analogs of the fields $\Pi_{ \bd{k} \parallel} ( \tau )$ and $\Pi_{
  \bd{k} \bot} ( \tau )$ of the NLSM.  The relation between these two
different parameterizations of the spin fluctuations is simply
\cite{Hasselmann05}
\begin{subequations}
  \begin{eqnarray}
    \hat{\Psi}_{ \bd{k} +} & = & -i
    ( \chi_0/2 V  E_{\bd{k} +})^{1/2}
    \left[ E_{\bd{k} +} \hat{\Pi}_{ \bd{k} +} 
      + i \chi_0^{-1} \hat{\Phi}_{ \bd{k} +} 
    \right], \hspace{8mm}
    \label{eq:dkplusmap}
    \\
    \hat{\Psi}_{ \bd{k} -} & = &( \chi_0/2 V  E_{\bd{k} -})^{1/2}
    \left[E_{\bd{k} -}   \hat{\Pi}_{ \bd{k} -}
      +i\chi_0^{-1} \hat{\Phi}_{ \bd{k} -}   
    \right]
    \label{eq:dkminusmap}
    \; , \hspace{7mm}
  \end{eqnarray}
\end{subequations}
where $V = N a^D$ is the volume of the system.
One easily verifies that the pairs $\hat{\Pi}_{\sigma}$, $\hat{\Phi}_{\sigma}$
 satisfy commutation relations of canonically
conjugate bosonic field operators,
\begin{eqnarray} 
  [   \hat{\Pi}_{ \bd{k}  \sigma} , 
  \hat{\Phi}_{ \bd{k}^{\prime} \sigma^{\prime} }]  =  i V 
  \delta_{ \bd{k} , - \bd{k}^{\prime} }  \delta_{ \sigma , \sigma^{\prime}}
  \; .
  \label{eq:cancom2}
\end{eqnarray}
The field operators $\hat{\Pi}_{ \bd{k} \sigma}$ and
$\hat{\Phi}_{ \bd{k} \sigma}$ 
 have a simple
interpretation in terms of transverse staggered and uniform spin
components. The
Fourier transformed spin operators on the $A$ and $B$
sublattice are given  
by 
  \begin{eqnarray}
    {\bd S}_{A/B,{\bd k}}&=&
(2/N)^{1/2} 
\sum_{\bd{r}_i \in A/B}
    e^{-i \bd{k}\cdot \bd{r}_i} \bd{S}_i 
    \; .
  \end{eqnarray}
Using Eqs.~(\ref{eq:spinexpansion},\ref{eq:transverse}) we 
find that, to leading order in $1/S$ and $m_0$,
the transverse staggered components of the
spins are given by\cite{footnote}
\begin{subequations}
  \begin{eqnarray}
    \hspace{-.7cm}
    S_{{\rm st},\bd{k}}^{(1)}&=&
    \frac{1}{\sqrt{2}}( S_{A,{\bd k}}^{(1)}-S_{B,{\bd k}}^{(1)})\approx 
    (S/a^D) N^{-1/2}
    \hat{\Pi}_{ \bd{k} -} \; , \\ 
    \hspace{-.7cm}
    S_{{\rm st},\bd{k}}^{(2)}&=&
    \frac{1}{\sqrt{2}}(
    S_{A,{\bd k}}^{(2)}+S_{B,{\bd k}}^{(2)})\approx
    -(S/a^D) N^{-1/2}
    \hat{\Pi}_{ \bd{k} +} . 
  \end{eqnarray}
  \label{staggcomps}
\end{subequations}

\noindent In contrast, the transverse components of the uniform
magnetization are
related to the  operators $\hat{\Phi}_{ \bd{k} \sigma}$,
\begin{subequations}
  \begin{eqnarray}
   S_{{\bd k}}^{(1)}= \frac{1}{\sqrt{2}}
   (S_{A,{\bd k}}^{(1)}+S_{B,{\bd k}}^{(1)})\approx 
   N^{-1/2}
   \hat{\Phi}_{ \bd{k} +} \; ,
   \\
   S_{{\bd k}}^{(2)}=\frac{1}{\sqrt{2}}
   (S_{A,{\bd k}}^{(2)}-S_{B,{\bd k}}^{(2)})\approx 
    N^{-1/2}
    \hat{\Phi}_{ \bd{k} -} \; .
  \end{eqnarray}
  \label{uniformcomps}
\end{subequations}

\noindent In terms of these operators, our quadratic spin-wave
Hamiltonian (\ref{eq:H2diagpsi}) can be written as
\begin{eqnarray}
  \hat{H}_2  =  \frac{1}{2 V} \sum_{ \bd{k},\sigma}
  \Bigl[
  \chi_0^{-1} \hat{\Phi}_{ - \bd{k} \sigma} \hat{\Phi}_{ \bd{k} \sigma} 
  + \chi_0
  E_{ \bd{k} \sigma}^2 \hat{\Pi}_{ - \bd{k} \sigma} \hat{\Pi}_{\bd{k} \sigma} 
  \Bigr]
  \; .
\end{eqnarray}

The effective Euclidean action $S_{\rm eff}[\Pi_\sigma]$ 
for the staggered spin fluctuations can
now be obtained by writing the partition function as a phase
space path integral over the fields $\Pi_{ \bd{k} \sigma} ( \tau )$
and $\Phi_{\bd{k} \sigma} ( \tau )$ associated with the above
operators and subsequently integrating over the 
$\Phi_{\sigma}$-fields which
represent
gapped ferromagnetic fluctuations.  At the level of a Gaussian
approximation, we obtain in this way $S_{\rm eff}\approx
S_{\rm eff}^{(2)}$ with
\begin{equation} 
  S_{\rm eff}^{(2)} [ \Pi_{\sigma} ]  = 
  \frac{ \chi_0}{2 \beta V }  \sum_{ K,\sigma } 
  (  \omega_n^2     +  E_{\bd{k} \sigma}^2  )
  \Pi_{ -K \sigma} \Pi_{ K \sigma}
  \; .
  \label{eq:P2Pires}
\end{equation}
Here, we have 
combined momenta $\bd{k}$ and bosonic Matsubara frequencies
$\omega_n$ in a composite label $K = ( \bd{k} , i \omega_n )$. We have further
defined
\begin{eqnarray}
\Pi_{ K \sigma }=\int_0^\beta d\tau  \; e^{  i \omega_n \tau }
 \Pi_{ \bd{k}  \sigma } ( \tau )
  \; .
  \label{eq:PiMatsubara}
\end{eqnarray}
At long wavelengths the effective action (\ref{eq:P2Pires}) has the
same form as the corresponding Gaussian part of the action of the
NLSM.  However, in contrast to the NLSM (\ref{eq:sigmah2}), our effective
action (\ref{eq:P2Pires}) has the correct spin-wave dispersion, even
for short wavelengths.

We have calculated the leading corrections to
the effective action $S_{\rm eff} [ \Pi_{\sigma} ]$ for staggered spin
fluctuations arising from spin-wave interactions in the
Holstein-Primakoff approach.\cite{Hasselmann05b} 
Keeping only interaction
terms which become relevant below three
dimensions and thus omitting terms which are marginal in $D=1$, we find
in the continuum limit 
\begin{eqnarray}
  S_{\rm eff} [ \Pi_{\sigma} ] &   = &  \frac{ \chi_0}{2 \beta V } 
  \sum_{ K,\sigma}
  \bigl( \omega_n^2 + c_{\sigma}^2 \bd{k}^2  + r_{ \sigma} \bigr) 
  \Pi_{ -K  \sigma}  \Pi_{ K  \sigma}
  \nonumber
  \\
  &  -  &  
  i   \chi_0   h   \int_0^{\beta} d \tau \int d^D r  \Pi_{+}^2 
  \partial_\tau \Pi_{ -}
  \nonumber
  \\
  & -  &    \frac{\chi_0 h^2 }{8}
  \int_0^{\beta} d \tau \int d^D r 
  \Pi_{+}^2 ( \Pi_{+}^2 + \Pi_{-}^2 )
  \; .
  \label{eq:Sefffinal1}
\end{eqnarray}
Here, the $c_{\sigma}$ are the Holstein-Primakoff results for
the spin-wave velocities, which for large $S$ 
are
given in Eq.~(\ref{eq:spinwaveh}).  
To leading
orders in $1/S$,
the values of the gap parameters are
$ r_{+}=h^2$ and $ r_{-}= 0$. For symmetry reasons, the true
spin-wave spectrum of the $\Pi_-$-mode
must remain gapless at vanishing wave-vector
(Goldstone mode),\cite{Golosov88} while the spin-wave gap of the $\Pi_+$-mode
is not renormalized\cite{Golosov88,Oshikawa02}.
In a renormalization group
analysis, this requires fine tuning such that at the fixed point
these conditions are met.
While the cubic term in Eq.~(\ref{eq:Sefffinal1})
corresponds precisely to the cubic (Berry phase) term in the NLSM
(\ref{eq:sigmah2}), the quartic interaction between the field
components is absent 
if the magnetic field is included in the NLSM
by the minimal coupling (\ref{eq:minimal}).

\vspace*{-6mm}

\section{Summary and conclusion}
\vspace{-3mm}
\label{sec:conclusions}

In this work we have used the Holstein-Primakoff transformation to
derive an effective action for  staggered transverse spin
fluctuations of quantum Heisenberg antiferromagnets in uniform
magnetic fields.  Our effective action contains an additional quartic
interaction between the field components which is not contained in the
NLSM.  It is easy to see that at zero temperature the quartic
interaction vertex in Eq.~(\ref{eq:Sefffinal1}) is relevant in the
renormalization group sense for dimensions $D < 3$, 
so that we expect that it
generates singularities in perturbation theory.\cite{Hasselmann05b}
We conclude that the NLSM given in Eq.~(\ref{eq:sigmah2}) does not
contain all relevant interactions in the ordered phase below three
dimensions.  The reason for this is that not all approximations that
are usually made in the derivation of the NLSM\cite{Sachdev99} 
are justified in the
presence of a uniform magnetic field.  In particular, it is not
justified to neglect Umklapp scattering processes between transverse
and longitudinal spin fluctuations involving momentum transfers across
the boundary of the magnetic Brillouin zone. 
As an alternative to Eq.~(\ref{eq:Sefffinal1}) it should be possible
to analyse the long-wavelength staggered spin fluctuations within a
phenomenological 
Ginzburg-Landau model with a quartic interaction term\cite{Affleck9091} 
since this model treats the longitudinal modes as  independent degrees
of freedom.

The authors thank Ian Affleck for illuminating discussions. 
This work was supported by the DFG via Forschergruppe FOR 412.


\begin{thebibliography}{99}
\bibitem{footnote}
We use here a basis which is rotated compared to the one used
in Ref.~[\onlinecite{Hasselmann05}]. This leads to different appearances
of the minus signs in the transformation Eqs.~(\ref{staggcomps},\ref{uniformcomps}).
\end{thebibliography}
\end{document}